\documentclass[runningheads]{llncs}
\usepackage[margin=1.3in]{geometry}
\usepackage{graphicx}
\usepackage{amsmath}
\usepackage{enumitem}
\usepackage{hyperref}
\usepackage{caption,subcaption}

\usepackage[ruled,vlined]{algorithm2e}
\DeclareMathOperator*{\argmax}{arg\,max}

\newcommand{\C}{\mathcal{C}}
\newcommand{\A}{\mathcal{A}}

\begin{document}
\title{Using synthetic networks for parameter tuning \\ in community detection}

\author{Liudmila Prokhorenkova}

\institute{Moscow Institute of Physics and Technology, Dolgoprudny, Russia \\
Yandex, Moscow, Russia \\
\email{ostroumova-la@yandex.ru}}

\maketitle              

\begin{abstract}
Community detection is one of the most important and challenging problems in network analysis. However, real-world networks may have very different structural properties and communities of various nature. As a result, it is hard (or even impossible) to develop one algorithm suitable for all datasets. A standard machine learning tool is to consider a parametric algorithm and choose its parameters based on the dataset at hand. However, this approach is not applicable to community detection since usually no labeled data is available for such parameter tuning. In this paper, we propose a simple and effective procedure allowing to tune hyperparameters of any given community detection algorithm without requiring any labeled data. The core idea is to generate a synthetic network with properties similar to a given real-world one, but with known communities. It turns out that tuning parameters on such synthetic graph also improves the quality for a given real-world network. To illustrate the effectiveness of the proposed algorithm, we show significant improvements obtained for several well-known parametric community detection algorithms on a variety of synthetic and real-world datasets. 

\keywords{Community detection \and Parameter tuning \and Hyperparameters \and LFR benchmark}
\end{abstract}

\section{Introduction} 

Community structure, which is one of the most important properties of complex networks, is characterized by the presence of groups of vertices (called communities or clusters) that are better connected to each other than to the rest of the network. In social networks, communities are formed based on common interests or on geographical location; on the Web, pages are clustered based on their topics; in protein-protein interaction networks, clusters are formed by proteins having the same specific function within the cell, and so on. Being able to identify communities is important for many applications: recommendations in social networks, graph compression, graph visualization, etc.

The problem of community detection has several peculiarities making it hard to formalize and, consequently, hard to develop a good solution for. First, as pointed out in several papers, there is no universal definition of communities~\cite{fortunato2016community}. As a result, there are no standard procedures for comparing the performance of different algorithms. Second, real-world networks may have very different structural properties and communities of various nature. Hence, it is impossible to develop one algorithm suitable for all datasets, as discussed in, e.g.,~\cite{peel2017ground}. A standard machine learning tool applied in such cases is to consider a parametric algorithm and tune its parameters based on the given dataset. Parameters which have to be chosen by the user based on the observed data are usually called \textit{hyperparameters} and are often tuned via cross-validation, but this procedure requires a training part of the datasets with available ground truth labels. However, the problem of community detection is unsupervised, i.e., no ground truth community assignments are given, so standard tuning approaches are not applicable and community detection algorithms are often non-parametric.

We present a surprisingly simple and effective method for tuning hyperparameters of any community detection algorithm which requires no labeled data and chooses suitable parameters based only on the structural properties of a given graph. The core idea is to generate a synthetic network with properties similar to a given real-world one, but with known community assignments, hence we can optimize the hyperparameters on this synthetic graph and then apply the obtained algorithm to the original real-world network. It turns out that such a trick significantly improves the performance of the initial algorithm. 

To demonstrate the effectiveness and the generalization ability of the proposed approach, we applied it to three different algorithms on various synthetic and real-world networks. In all cases, we obtained substantial improvements compared to the algorithms with default parameters. However, since communities in real-world networks cannot be formally defined, it is impossible to provide any theoretical guarantees for those parameter tuning strategies which do not use labeled data. As a result, the quality of any parameter tuning algorithm can be demonstrated only empirically. Based on the excellent empirical results obtained, we believe that the proposed approach captures some intrinsic properties of real-world communities and would generalize to other datasets and algorithms.

\section{Background and related work}\label{sec:related_work}

During the past few years, many community detection algorithms have been proposed, see~\cite{coscia2011classification,fortunato2010community,fortunato2016community,malliaros2013clustering} for an overview. In this section, we take a closer look at the algorithms and concepts used in the current research. 

\subsection{Modularity}\label{sec:modularity}

Let us start with some notation. We are given a graph $G = (V,E)$, $V$ is a set of $n$ vertices, $E$ is a set of $m$ undirected edges. Denote by $\C$ a partition of $V$ into several disjoint communities: $\C = \{C_1, \ldots, C_k\}$. Also, let $m_{in}$ and $m_{out}$ be the number of intra- and inter-cluster edges in a graph $G$ partitioned according $\C$. Finally, $d(i)$ denotes the degree of a vertex $i$ and $D(C) = \sum_{i \in C} d(i)$ is the overall degree of a community $C \in \C$.

\textit{Modularity} is a widely used measure optimized by many community detection algorithms. It was first proposed in~\cite{newman2004finding} and is defined as follows
\begin{equation}\label{eq:modularity}
Q(\C,G,\gamma) = \frac{m_{in}}{m} -  \frac{\gamma}{4 m^2} \sum_{C \in \C} D(C)^2\,,
\end{equation}
where $\gamma$ is a resolution parameter~\cite{lancichinetti2011limits}. The intuition behind modularity is the following: the first term in~\eqref{eq:modularity} is the fraction of intra-cluster edges, which is expected to be relatively high for good partitions, while the second term penalizes this value for having too large communities. Namely, the value $\frac{\sum_{C \in \C} D(C)^2}{{4 m^2}}$ is the expected fraction of intra-cluster edges if we preserve the degree sequence but connect all vertices randomly, i.e., if we assume that our graph is constructed according to the configuration model~\cite{molloy1995critical}.

Modularity was originally introduced with $\gamma = 1$ and many community detection algorithms maximizing this measure were proposed. However, it was shown in~\cite{fortunato2007resolution} that modularity has a resolution limit, i.e., algorithms based on modularity maximization are unable to detect communities smaller than some size. Adding a resolution parameter allows to overcome this problem: larger values of $\gamma$ in general lead to smaller communities. However, tuning $\gamma$ is a  challenging task. In this paper, we propose a solution to this problem.

\subsection{Modularity optimization and Louvain algorithm}\label{sec:louvain}

Many community detection algorithms are based on modularity optimization.
In this paper, as one of our base algorithms, we choose arguably the most well-known and widely used greedy algorithm called Louvain~\cite{blondel2008fast}. It starts with each vertex forming its own community and works in several phases. To create the first level of a partition, we iterate through all vertices and for each vertex $v$ we compute the gain in modularity coming from removing $v$ from its community and putting it to each of its neighboring communities; then we move $v$ to the community with the largest gain, if it is positive. When we cannot improve modularity by such local moves, the first level is formed. After that, we replace the obtained communities by supervertices connected by weighted edges; the weight between two supervertices is equal to the number of edges between the vertices of the corresponding communities. Then we repeat the process described above with the supervertices and form the second level of a partition. After that, we merge the supervertices again, and so on, as long as modularity increases. The Louvain algorithm is quite popular since it is fast and was shown to provide partitions of good quality. However, by default, it optimizes modularity with $\gamma = 1$, therefore, it suffers from a resolution limit. 

\subsection{Likelihood optimization methods}\label{sec:likelihood_optimization}

Likelihood optimization algorithms are also widely used in community detection. Such methods are mathematically sound and have theoretical guarantees under some model assumptions~\cite{bickel2009nonparametric}. The main idea is to assume some underlying random graph model parameterized by community assignments and find a partition $\C$ that maximizes the likelihood $P(G|\C)$, which is the probability that a graph generated according to the model with communities $\C$ exactly equals $G$. 

The standard random graph model assumed by likelihood maximization methods is the stochastic block model (SBM) or its simplified version~--- planted partition model (PPM). In these models, the probability that two vertices are connected by an edge depends only on their community assignments. Recently, the degree-corrected stochastic block model (DCSBM) together with the degree-corrected planted partition model (DCPPM) were proposed~\cite{karrer2011stochastic}. These models take into account the observed degree sequence of a graph, and, as a result, they are more realistic. It was also noticed that if we fix the parameters of DCPPM, then likelihood maximization based on this model is equivalent to modularity optimization with some $\gamma$~\cite{newman2016community}. Finally, in a recent paper~\cite{prokhorenkova2018community} the independent LFR model (ILFR) was proposed and analyzed. It was shown that ILFR gives a better fit for a variety of real-world networks~\cite{prokhorenkova2018community}. In this paper, to illustrate the generalization ability of the proposed hyperparameter tuning strategy, in addition to the Louvain algorithm, we also use parametric likelihood maximization methods based on PPM and ILFR.

\subsection{LFR model}\label{sec:lfr}

Our parameter tuning strategy is based on constructing a synthetic graph structurally similar to the observed network. To do this, we use the LFR model~\cite{lancichinetti2008benchmark} which is the well-known synthetic benchmark usually used for comparison of community detection algorithms. LFR generates a graph with power-law distributions of both degrees and community sizes in the following way. First, we generate the degrees of vertices by sampling them independently from the power-law distribution with exponent $\gamma_d$, mean $\bar d$ and with maximum degree $d_{max}$. Then, using a mixing parameter $\hat \mu$, $0 < \hat\mu < 1$, we obtain internal and external degrees of vertices: we expect each vertex to share a fraction $1-\hat\mu$ of its edges with the vertices of its community and a fraction $\hat\mu$ with the other vertices of the network. After that, the sizes of the communities are sampled from a power-law distribution with exponent $\gamma_C$ and minimum and maximum community sizes $C_{min}$ and $C_{max}$, respectively. Then, vertices are assigned to communities such that the internal degree of any vertex is less than the size of its community. Finally, the configuration model~\cite{molloy1995critical} with rewiring steps is used to construct a graph with a given degree sequence and with the required fraction of internal edges. The detailed description of this procedure can be found in~\cite{lancichinetti2008benchmark}.

\section{Tuning parameters}\label{sec:tuning_parameters}

Assume that we are given a graph $G$ and our aim is to find a partition $\C$ of its vertex set into disjoint communities. To do this, we have a community detection algorithm $\A_{\theta}$, where $\theta\in \Theta$ is a set of hyperparameters. Let $\theta_0$ be the default hyperparameters. Assume that we are also given a quality function $Q(\C_{\A_{\theta}},\C_{GT})$ allowing to measure goodness of a partition $\C_{\A_{\theta}}$ obtained by $\A_{\theta}$ compared to the ground truth partition $\C_{GT}$. Ideally, we would like to find $\bar\theta = \argmax_\theta Q(\C_{\A_{\theta}},\C_{GT})$. However, we cannot do this since $\C_{GT}$ is not available. Therefore, we propose to construct a synthetic graph $G'$ which has structural properties similar to $G$ and also has known community assignments. For this purpose, we use the LFR model described in Section~\ref{sec:lfr}. To apply this model, we have to define its parameters, which can be divided into \textit{graph-based} ($n$, $\gamma_d$, $\bar d$, $d_{max}$) and \textit{community-based} ($\gamma_C$, $C_{min}$, $C_{max}$, $\hat \mu$). 

Graph-based parameters are easy to estimate:
\begin{itemize}
\item $n = |V(G)|$ is the number of vertices in the observed network;
\item $\bar d = \frac{2|E(G)|}{n}$ is the average degree;
\item $d_{max}$ is the maximum degree in $G$;
\item $\gamma_d$ is the exponent of the power-law degree distribution; we estimate this parameter by fitting the power-law distribution to the cumulative degree distribution (we minimize the sum of the squared residuals in log-log scale).
\end{itemize}

Community-based parameters contain some information about the community structure, which is not known for the graph $G$. However, we can try to approximate these parameters by applying the algorithm $\A_{\theta_0}$ with default parameters to $G$. This would give us some partition $\C_0$ which can be used to estimate the remaining parameters:
\begin{itemize}
\item $\hat \mu = \frac{m_{out}}{m}$ is the mixing parameter, i.e., the fraction of inter-community edges in $G$ partitioned according to $\C_0$;
\item $\gamma_C$ is the exponent of the power-law community size distribution; we estimate this parameter by fitting the power-law distribution to the cumulative community size distribution obtained from $\C_0$ (we minimize the sum of the squared residuals in log-log scale);
\item $C_{min}$ and $C_{max}$ are the minimum and maximum community sizes in $\C_0$.
\end{itemize}

We generate a graph $G'$ according to the LFR model with parameters specified above. Using $G'$ we can tune the parameters to get a better value of $\theta$:
\begin{equation}\label{eq:opt_theta}
\theta_{opt} = \argmax_\theta Q(\C'_{\A_{\theta}},\C_{GT}')\,,
\end{equation}
where $\C_{GT}'$ is known ground truth partition for $G'$ and $\C'_{\A_{\theta}}$ is a partition of $G'$ obtained by $\A_{\theta}$. It turns out that this simple idea leads to a universal method for tuning $\theta$, which successfully improves the results of several algorithms $\A_{\theta}$ on a variety of synthetic and real-world datasets, as we show in Section~\ref{sec:experiments}.

\begin{algorithm}
\SetKwInOut{Input}{input}\SetKwInOut{Output}{output}
	\Input{\,\,\,graph $G$, algorithm $\A_{\theta}$, default hyperparameters $\theta_0$, candidate parameters $\{\theta_i\}_{i=1}^l$, quality function $Q$, $n_{graphs}$, $n_{runs}$ }
    \BlankLine
    $n, \bar d, d_{max}, \gamma_{d} \leftarrow EstimateGraphParams(G)$\; 
    $\C_0 \leftarrow \A_{\theta_0}(G)$\;
    $\hat\mu,\gamma_C, C_{min}, C_{max} \leftarrow EstimateCommunityParams(G,\C_0)$\;
    $ParamsList \leftarrow \emptyset$\;
    \For{$i \leftarrow 1$ \KwTo $n_{graphs}$}{
    $G', \C_{GT}' \leftarrow GenerateLFR(n, \bar d, d_{max}, \gamma_{d}, \hat\mu,\gamma_C, C_{min}, C_{max})$\;
    $QualityList \leftarrow \emptyset$\;
    \For{$\theta \in \{\theta_i\}_{i=1}^l$}{
    $Qualities \leftarrow \emptyset$\; 
    \For{$j \leftarrow 1$ \KwTo $n_{runs}$}{
    $\C_{\theta} \leftarrow \A_{\theta}(G')$\;
    Add $Q(\C_{\theta},\C_{GT}')$ to $Qualities$\;
    }
    $MeanQuality \leftarrow \mathrm{mean}(Qualities)$\;
    Add $MeanQuality$ to $QualityList$\;
    } 
    $index \leftarrow \argmax(QualityList)$\;
    Add $\theta_{index}$ to $ParamsList$\;
     }
    $\theta = median(ParamsList)$\;
    \Return{$\theta$}
	{\caption{Hyperparameter tuning}
    \label{alg:tuning}}
\end{algorithm}

The detailed description of the proposed procedure is given in Algorithm~\ref{alg:tuning}. Note that in addition to the general idea described above we also propose two modifications improving the robustness of the algorithm. The first one reduces the effect of randomness in the LFR benchmark: if the number of vertices in $G$ is small, then a network generated by the LFR model can be noisy and the optimal parameters $\theta_{opt}$ computed according to Equation~\eqref{eq:opt_theta} may vary from sample to sample. Hence, we propose to generate $n_{graphs}$ synthetic networks and take the median of the obtained parameters. The value $n_{graphs}$ depends on computational resources: larger values, obviously, lead to more stable results. Fortunately, as we discuss in Section~\ref{sec:stability}, this effect of randomness is critical only for small graphs, so we do not have to increase computational complexity much for large datasets. 

The second improvement accounts for a possible randomness of the algorithm $\A_{\theta}$. If $\A_{\theta}$ includes some random steps, then we can increase the robustness of our procedure by running $\A_{\theta}$ several times and averaging the obtained qualities. The corresponding parameter is called $n_{runs}$ in Algorithm~\ref{alg:tuning}. Formally, in this case Equation~\eqref{eq:opt_theta} should be replaced by 
\begin{equation}\label{eq:opt_theta_honest}
\theta_{opt} = \argmax_\theta \frac{1}{n_{runs}} \sum_{i=1}^{n_{runs}} Q(\C'_{\A_{\theta},i},\C_{GT}')\,,
\end{equation}
where $\C'_{\A_{\theta},i}$ is a (random) partition obtained by $\A_{\theta}$ on $G'$. If $\A_{\theta}$ is deterministic, then it is sufficient to take $n_{runs} = 1$.

Note that for the sake of simplicity in Algorithm~\ref{alg:tuning} we use grid search to approximately find $\theta_{opt}$ defined in~\eqref{eq:opt_theta_honest}. However, any other method of black-box optimization can be used instead, e.g., random search~\cite{bergstra2012random}, Bayesian optimization~\cite{snoek2015scalable}, Gaussian processes~\cite{Golovin2017Googlevizier}, sequential model-based optimization~\cite{hutter2011sequential}, and so on. More advanced black-box optimization methods can significantly speed up the algorithm. 

Let us discuss the time complexity of the proposed algorithm. If complexity of $\A_{\theta}$ is $f(G)$, then complexity of Algorithm~\ref{alg:tuning} is $O\left(f(G)\cdot l\cdot n_{runs}\cdot n_{graphs}\right)$, where $l$ is the number of steps made by the black-box optimization (the complexity of generating $G'$ is usually negligible compared with community detection). In other words, the complexity is $n_{runs}\cdot n_{graphs}$ times larger than the complexity of any black-box parameter optimization algorithm. However, as we discuss in Section~\ref{sec:stability}, $n_{runs}$ and $n_{graphs}$ can be equal to one for large datasets.

Finally, note that it can be tempting to make several iterations of Algorithm~\ref{alg:tuning} to further improve $\theta_{opt}$. Namely, in Algorithm~\ref{alg:tuning} we estimate community-based parameters of LFR using the partition $\C_0$ obtained with $\A_{\theta_0}$. Then, we obtain better parameters $\theta_{opt}$. These parameters can be further used to get a better partition using $\A_{\theta_{opt}}$ and this partition is expected to give even better community-based parameters. However, in our preliminary experiments, we did not notice significant improvements from using several iterations, therefore we propose to use Algorithm~\ref{alg:tuning} as it is without increasing its computational complexity. 

\section{Experiments}\label{sec:experiments}

\subsection{Parametric algorithms}\label{sec:algorithms}

We use the following algorithms to illustrate the effectiveness of the proposed hyperparameter tuning strategy.

\paragraph{Louvain}
This algorithm is described in Section~\ref{sec:louvain}, it has the resolution parameter $\gamma$ with default value $\gamma_0 = 1$. We take the publicly available implementation from~\cite{prokhorenkova2018community},\footnote{\url{https://github.com/altsoph/community_loglike}} where the algorithm is called DCPPM since modularity maximization is equivalent to the likelihood optimization for the DCPPM random graph model.

\paragraph{PPM}
This algorithms is based on likelihood optimization for PPM (see Section~\ref{sec:likelihood_optimization}). We use the publicly available implementation proposed in~\cite{prokhorenkova2018community}, where the Louvain algorithm is used as a basis to optimize the likelihood for several models. Since likelihood optimization for PPM is equivalent to maximizing a simplified version of modularity based on the Erd{\H{o}}s--R{\'e}nyi model instead of the configuration model~\cite{newman2016community}, PPM algorithm also has a resolution parameter $\gamma$ with the default value $\gamma_0 = 1$. 

\paragraph{ILFR}
This is a likelihood optimization algorithm based on the ILFR model (see Section~\ref{sec:likelihood_optimization}). Again, we use the publicly available implementation from~\cite{prokhorenkova2018community}. ILFR algorithm has one parameter $\mu$ called mixing parameter and no default value for this parameter is proposed in the literature. In this paper, we take $\mu_0 = 0.3$, which is close to the average mixing parameter in the real-world datasets under consideration (see Section~\ref{sec:datasets}). Our experiments confirm that $\mu_0 = 0.3$ is a reasonable default value for this algorithm.

\vspace{5pt}

Let us stress that in this paper we are not aiming to develop the best community detection algorithm or to analyze all existing methods. Our main goal is to show that hyperparameter tuning is possible in the field of community detection. We use several base algorithms described above to illustrate the generalization ability of the proposed approach. For each algorithm, our aim is to improve its default parameter by our parameter tuning strategy.

\subsection{Datasets}\label{sec:datasets}

\paragraph{Synthetic networks}

We generated several synthetic graphs according to the LFR benchmark described in Section~\ref{sec:lfr} with $n = 10^4$, $\gamma_d = 2.5$, $\bar d = 20$, $d_{max} = 200$, $\gamma_C = 1.5$, $C_{min} = 50$, $C_{max} = 500$, $\hat\mu \in \{0.4,0.5,0.6,0.7\}$.\footnote{Note that $\hat\mu>0.5$ does not mean the absence of community structure since usually a community is much smaller than the rest of the network and even if more than a half of the edges for each vertex go outside the community, the density of edges inside the community is still large.}
On the one hand, one would expect results obtained on such synthetic datasets to be optimistic, since the same LFR model is used both to tune the parameters and to validate the performance of the algorithms. On the other hand, recall that the most important ingredient of the model, i.e., the distribution of community sizes, is not known and has to be estimated using the initial community detection algorithm, and incorrect estimates may negatively affect the final performance. 

\paragraph{Real-world networks}

We follow the work~\cite{prokhorenkova2018community}, where the authors collected and shared 8 real-world datasets publicly available in different sources.\footnote{\url{https://github.com/altsoph/community_loglike/tree/master/datasets}} For all these datasets, the ground truth community assignments are available and the communities are non-overlapping. These networks are of various sizes and structural properties, see the description in Table~\ref{tab:datasets}.

\begin{table}[t]
  \caption{Real-world datasets}
  \label{tab:datasets}
  \centering
  \begin{tabular}{lcccc}
    \hline
Dataset& \hspace{15pt}$n$\hspace{15pt} & $m$ & \,\,num. clusters\,\, & mixing parameter\\
    \hline
    Karate club~\cite{zachary1977information} 
    	& 34 & 78 & 2 & 0.128 \\
    Dolphin network~\cite{lusseau2003bottlenose} 
    	& 62 & 159 & 2 & 0.038  \\
	College football~\cite{newman2004finding} 
    	& 115 & 613 & 11 & 0.325 \\
        	Political books~\cite{newman2006modularity} & 105 & 441 & 3 & 0.159 \\
	Political blogs~\cite{adamic2005political} 
    	& 1224 & 16715 & 2 & 0.094 \\
    email-Eu-core~\cite{leskovec2007graph} & 986 & 16064 & 42 &  0.664  \\
    Cora citation~\cite{vsubelj2013model} & 24166 & 89157 & 70 & 0.458 \\
AS~\cite{boguna2010sustaining} & 23752 & 58416 &   176 & 0.561 \\
  \hline
\end{tabular}
\end{table}

\subsection{Evaluation metrics}\label{sec:metrics}

In the literature, there is no universally accepted metric for evaluating the performance of community detection algorithms. Therefore, we analyze several standard ones~\cite{fortunato2010community}. Namely, we use two widely used similarity measures based on counting correctly and incorrectly classified pairs of vertices: Rand and Jaccard indices. We also consider the Normalized Mutual Information (NMI) of two partitions: if NMI is close to 1, one needs a small amount of information to infer the ground truth partition from the obtained one, i.e., two partitions are similar.

\subsection{Experimental setup}\label{sec:setup}

We apply the proposed strategy to the algorithms described in Section~\ref{sec:algorithms}. We use the grid search to find the parameter $\theta_{opt}$ (we do this to make our results easier to reproduce and we also need this for the analysis of stability in Section~\ref{sec:stability}). For ILFR we try $\mu$ in the range $[0,1]$ with step size 0.05 and for Louvain and PPM on real-world datasets we take $\gamma$ in the range $[0,2]$ with step size $0.1$. Although we noticed that in some cases the optimal $\gamma$ for PPM and Louvain can be larger than 2, such cases rarely occur on real-world datasets. On synthetic graphs, we take $\gamma$ in the range $[0,4]$ (with step size 0.2) to demonstrate the behavior of $\gamma_{opt}$ depending on~$\hat \mu$.

To guarantee stability and reproducibility of the obtained results, we choose a sufficiently large parameter $n_{runs}$, although we noticed similar improvements with much smaller values. Namely, for Karate, Dolphins, Football, and Political books we take $n_{runs} = 10^3$, for Political blogs and Eu-core  $n_{runs} = 100$, for Cora, AS, and synthetic networks $n_{runs} = 2$. We take $n_{graphs} = 10^3$ for four smallest datasets and $n_{graphs} = 100$ for the other ones (we choose such large values to plot the histograms on Figure~\ref{fig:hist}).

Finally, note that it is impossible to measure the statistical significance of obtained improvements on real-world datasets since we have only one copy for each graph. However, we can account for the randomness included in the algorithms. Namely, Louvain, PPM, and ILFR are randomized, since at each iteration they order the vertices randomly. Therefore, to measure if $\theta_{opt}$ is significantly better or worse than $\theta_0$, we can run each algorithm several times and then apply the unpaired t-test (we use 100 runs in all cases).

\subsection{Results}

In this section, we first discuss the improvements obtained for each algorithm and then analyze the stability of the parameter tuning strategy and the effect of the parameter $n_{graphs}$.

\subsubsection{Louvain algorithm}

In Table~\ref{tab:louvain}, for each similarity measure we present the value for the baseline algorithm (with $\gamma = 1$), the value for the tuned algorithm, and the obtained parameter $\gamma_{opt}$. Since Louvain is randomized, we provide the mean value together with an estimate of the standard deviation, which is given in brackets. The number of runs used to compute these values depends on the size of the dataset and on the available computational resources:
$10^4$ for Karate, Dolphins, Football and Political books, $10^3$ for Political blogs and Eu-core, 100 for Cora, AS and synthetic datasets.

\begin{table*}[t]
	\caption{Louvain algorithm, default value is $\gamma_0 = 1$, standard deviation is given in the brackets}
	\label{tab:louvain}
	\centering
	\begin{scriptsize}
		\begin{tabular}{l|cc|cc|c|cc|cc|c|cc|cc|c}
			\multicolumn{1}{c}{} & \multicolumn{5}{c|}{Rand} & \multicolumn{5}{c|}{Jaccard} & \multicolumn{5}{c}{NMI}\\
			Dataset & 
			\multicolumn{2}{c}{Default} &
			\multicolumn{2}{c}{Tuned} & $\gamma_{opt}$ & 
			\multicolumn{2}{c}{Default} & \multicolumn{2}{c}{Tuned} & $\gamma_{opt}$ & 
			\multicolumn{2}{c}{Default}
			& \multicolumn{2}{c}{Tuned} & $\gamma_{opt}$ \\
			\hline
			\hline
			Karate & 
			0.761 & (0.024) & \textbf{0.945} & (0.018) & 0.6 &
			0.520 & (0.042) & \textbf{0.892} & (0.030) & 0.5 &
			0.634 & (0.051) & \textbf{0.739} & (0.067) & 0.7 \\
			Dolphins &
			0.648 & (0.021) & \textbf{0.873} & (0.069) & 0.5 &
			0.374 & (0.037) & \textbf{0.608} & (0.133) & 0.1 &
			0.515 & (0.039) & 0.515 & (0.039) & 1.0 \\
			Football &
			0.970 & (0.010) & \textbf{0.992} & (0.004) & 1.7 &
			0.722 & (0.063) & \textbf{0.903} & (0.036) & 1.7 &
			0.923 & (0.016) & \textbf{0.969} & (0.008) & 1.7 \\
			Political books &
			0.828 & (0.024) & \textbf{0.845} & (0.005) & 0.8 &
			0.609 & (0.055) & \textbf{0.654} & (0.009) & 0.8 &
			0.542 & (0.024) & \textbf{0.560} & (0.011) & 0.8 \\
			Political blogs &
			0.883 & (0.004) & \textbf{0.901} & (0.001) & 0.7 &
			0.782 & (0.006) & \textbf{0.818} & (0.001) & 0.7 &
			0.635 & (0.007) & \textbf{0.678} & (0.007) & 0.8 \\
			Eu-core &
			0.862 & (0.020) & \textbf{0.932} & (0.004) & 1.4 &
			0.217 & (0.022) & \textbf{0.348} & (0.014) & 1.4 &
			0.576 & (0.018) & \textbf{0.656} & (0.009) & 1.4 \\
			Cora &
			0.941 & (0.002) & \textbf{0.964} & (0.001) & 2.0 &
			0.125 & (0.005) & \textbf{0.146} & (0.004) & 2.0 &
			0.457 & (0.005) & \textbf{0.494} & (0.004) & 2.0 \\
			AS &
			0.819 & (0.003) & \textbf{0.823} & (0.001) & 1.8 &
			0.190 & (0.026) & \textbf{0.258} & (0.013) & 0.6 &
			0.488 & (0.007) & 0.489 & (0.010) & 0.8 \\
			\hline
			\hline
			LFR-0.4 &
			0.999 & (0.001) & \textbf{1.000} & (0.000) & 2.8 &
			0.965 & (0.037) & \textbf{1.000} & (0.000) & 2.8 &
			0.994 & (0.003) & \textbf{1.000} & (0.000) & 2.8 \\
			LFR-0.5 &
			0.996 & (0.002) & \textbf{1.000} & (0.000) & 3.0 &
			0.861 & (0.078) & \textbf{0.997} & (0.007) & 3.0 &
			0.981 & (0.007) & \textbf{1.000} & (0.001) & 3.0 \\
			LFR-0.6 &
			0.984 & (0.008) & \textbf{0.999} & (0.000) & 3.6 &
			0.614 & (0.117) & \textbf{0.971} & (0.010) & 3.6 &
			0.940 & (0.020) & \textbf{0.992} & (0.002) & 3.6 \\
			LFR-0.7 &
			0.911 & (0.014) & \textbf{0.978} & (0.001) & 3.8 &
			0.089 & (0.024) & \textbf{0.320} & (0.032) & 3.6 &
			0.388 & (0.051) & \textbf{0.678} & (0.024) & 3.8 \\
			\hline
		\end{tabular}
	\end{scriptsize}
\end{table*}

One can see that our tuning strategy improves (or does not change) the results in all cases and the obtained improvements can be huge. For example, on Karate we obtain remarkable improvements from $0.761$ to $0.945$ (relative change is $24\%$) according to Rand and from $0.52$ to $0.892$ ($72\%$) according to Jaccard; on Dolphins we get $35\%$ improvement for Rand and $63\%$ for Jaccard; on Football we obtain plus $25\%$ for Jaccard; and so on. 
As discussed in Section~\ref{sec:setup}, we measured the statistical significance of the obtained improvements. The results which are significantly better are marked in bold in Table~\ref{tab:louvain}. On real-world datasets, all improvements except the one for NMI on AS are statistically significant (p-value $\ll 0.01$).\footnote{The results in Tables~\ref{tab:louvain}-\ref{tab:ilfr} are rounded to three decimals, so there may be a statistically significant improvement even when the numbers in the table are equal. Also, standard deviation less than 0.0005 is rounded to zero.} Let us note that in many cases the results of the tuned algorithm are much better than the best results reported in~\cite{prokhorenkova2018community}, where the authors used other strategies for choosing the hyperparameter values.\footnote{For small datasets, our results for the default Louvain algorithm may differ from the ones reported in~\cite{prokhorenkova2018community}. The reason is the high values of standard deviation. The authors of~\cite{prokhorenkova2018community} averaged the results over 5 runs of the algorithm, while we use more runs, i.e., our average values are more stable.}

For synthetic datasets, we also observe huge improvements and all of them are statistically significant. While for $\hat\mu \in \{0.4, 0.5\}$ the default algorithm can be considered as good enough, for large values of $\hat \mu$, $\hat\mu \in \{0.6, 0.7\}$, the tuned one is much better. For example, for LFR-0.7 the tuned parameter gives Jaccard index almost 4 times larger than the default one.

We noticed that for most of the datasets the values of $\gamma_{opt}$ computed using different similarity measures are the same or close to each other. However, there are some exceptions. The first one is Dolphins, where for Jaccard $\gamma_{opt} = 0.1$, for Rand $\gamma_{opt} = 0.5$, for NMI $\gamma_{opt} = 1.0$. We checked that if we take the median value $\gamma_{opt} = 0.5$, then for all measures we obtain statistically significant improvements, which seems to be another way to increase the stability of our strategy. 
The most notable case, where $\gamma_{opt}$ significantly differs for different similarity measures, is AS dataset, where $\gamma_{opt} = 1.8 > \gamma_0$ for Rand, $\gamma_{opt} = 0.6 < \gamma_0$ for Jaccard, and $\gamma_{opt} = 0.8 < \gamma_0$ for NMI. We will further make similar observations for other algorithms on this dataset. Such instability may mean that this dataset does not have a clear community structure (which can sometimes be the case for real-world networks~\cite{miasnikof2019statistical}).

\begin{table}[t]
	\caption{PPM algorithm, default value $\gamma_0 = 1$, standard deviation is given in the brackets}
	\label{tab:ppm}
	\centering
	\begin{scriptsize}
		\begin{tabular}{l|cc|cc|c|cc|cc|c|cc|cc|c}
			\multicolumn{1}{c}{} & \multicolumn{5}{c|}{Rand} & \multicolumn{5}{c|}{Jaccard} & \multicolumn{5}{c}{NMI}\\
			Dataset & 
			\multicolumn{2}{c}{Default} &
			\multicolumn{2}{c}{Tuned} & $\gamma_{opt}$ & 
			\multicolumn{2}{c}{Default} & \multicolumn{2}{c}{Tuned} & $\gamma_{opt}$ & 
			\multicolumn{2}{c}{Default}
			& \multicolumn{2}{c}{Tuned} & $\gamma_{opt}$ \\
			\hline
			\hline
			Karate & 
			0.756 & (0.024) & \textbf{0.782} & (0.041) & 0.8 &
			\textbf{0.509} & (0.040) & 0.487 & (0.000) & 0.1 &
			0.629 & (0.050) & 0.628 & (0.049) & 1.0 \\
			Dolphins &
			0.622 & (0.025) & \textbf{0.761} & (0.043) & 0.7 &
			0.330 & (0.042) & \textbf{0.815} & (0.189) & 0.1 &
			\textbf{0.466} & (0.045) & 0.411 & (0.024) & 1.6 \\
			Football &
			0.969 & (0.007) & \textbf{0.992} & (0.004) & 1.6 &
			0.716 & (0.041) & \textbf{0.901} & (0.040) & 1.6 &
			0.923 & (0.011) & \textbf{0.969} & (0.008) & 1.6 \\
			Political books &
			0.780 & (0.016) & \textbf{0.845} & (0.008) & 0.7 &
			0.481 & (0.038) & \textbf{0.647} & (0.016) & 0.7 &
			0.498 & (0.015) & \textbf{0.566} & (0.014) & 0.7 \\
			Political blogs &
			0.649 & (0.025) & \textbf{0.724} & (0.039) & 0.4 &
			0.315 & (0.022) & \textbf{0.471} & (0.037) & 0.4 &
			0.287 & (0.025) & \textbf{0.328} & (0.033) & 0.6 \\
			Eu-core &
			\textbf{0.800} & (0.021) & 0.774 & (0.024) & 0.9 &
			\textbf{0.099} & (0.012) & 0.091 & (0.011) & 0.9 &
			\textbf{0.529} & (0.018) & 0.490 & (0.016) & 0.8 \\
			Cora & 
			0.936 & (0.003) & \textbf{0.959} & (0.001) & 2.0 &
			0.115 & (0.004) & \textbf{0.130} & (0.004) & 2.0 &
			0.470 & (0.005) & \textbf{0.500} & (0.003) & 2.0 \\
			AS &
			0.793 & (0.012) & \textbf{0.815} & (0.004) & 1.8 &
			0.113 & (0.013) & \textbf{0.152} & (0.031) & 0.8 &
			0.459 & (0.020) & 0.459 & (0.018) & 1.2 \\
			\hline
			\hline
			LFR-0.4 &
			1.000 & (0.001) & \textbf{1.000} & (0.000) & 2.8 &
			0.987 & (0.024) & \textbf{0.995} & (0.020) & 2.8 &
			0.998 & (0.002) & \textbf{1.000} & (0.001) & 2.8 \\
			LFR-0.5 &
			0.996 & (0.005) & \textbf{0.999} & (0.001) & 3.0 &
			0.877 & (0.133) & \textbf{0.961} & (0.053) & 3.0 &
			0.989 & (0.013) & \textbf{0.995} & (0.007) & 3.0 \\
			LFR-0.6 &
			0.966 & (0.024) & \textbf{0.991} & (0.005) & 3.2 &
			0.438 & (0.201) & \textbf{0.667} & (0.090) & 3.2 &
			0.847 & (0.107) & \textbf{0.911} & (0.043) & 3.0 \\
			LFR-0.7 &
			0.801 & (0.027) & \textbf{0.966} & (0.009) & 2.8 &
			0.026 & (0.021) & \textbf{0.148} & (0.070) & 2.8 &
			0.181 & (0.109) & \textbf{0.510} & (0.109) & 2.8 \\
			\hline
		\end{tabular}
	\end{scriptsize}
\end{table}

\subsubsection{PPM algorithm}

For PPM (Table~\ref{tab:ppm}), our strategy improves the original algorithm for all real-world datasets but Eu-core (for all similarity measures), Karate (only for Jaccard), and Dolphins (only for NMI). Note that Karate and Dolphins are the only datasets (except for AS, which will be discussed further in this section), where the obtained values for $\gamma_{opt}$ are quite different for different similarity measures. We checked that if for these two datasets we take the median value of $\gamma_{opt}$, (0.8 for Karate and 0.7 for Dolphins), then we obtain improvements in all six cases, five of them, except NMI on Karate, are statistically significant (p-value $\ll 0.01$). On Eu-core the quality of PPM with $\gamma_0 = 1$ is worse than the quality of Louvain with $\gamma = 1$. This seems to be the reason why PPM chooses a suboptimal parameter $\gamma_{opt}$: a partition obtained by PPM does not allow for a good estimate of the community-based parameters.  As for Louvain, in many cases the obtained improvements are huge: e.g., the relative improvement for the Jaccard index is 147\% on Dolphins, 26\% on Football, 35\% on Political books, 50\% on Political blogs, an so on. All improvements are statistically significant.

We also improve the default algorithm on all synthetic datasets and for all similarity measures. As for the Louvain algorithm, the improvements are especially huge for large $\hat \mu$, $\hat \mu \in \{0.6,0.7\}$. All improvements are statistically significant.

\subsubsection{ILFR algorithm}\label{sec:exp_ilfrs}

For real-world datasets, in almost all cases, we obtain significant improvements (see Table~\ref{tab:ilfr}). One exception is Dolphins for NMI. This, again, can be fixed by taking a median of the values $\mu_{opt}$ obtained for all similarity measures on this dataset: $\mu_{opt} = 0.15$ improves the results compared to $\mu_0 = 0.3$ for all three measures. Other bad examples are Cora and AS, where Rand and NMI decrease, while Jaccard increases. For all other datasets, we obtain improvements. In many cases, the difference is huge and statistically significant.
On synthetic datasets, the default ILFR algorithm is the best among the considered ones.
In some cases, however, the default algorithm is further improved by our hyperparameter tuning strategy, while in others the difference is not statistically significant. Surprisingly, for large values of $\hat \mu$ the tuned value $\mu_{opt}$ is much smaller than $\hat \mu$. For example, for $\hat \mu = 0.6$ we get $\mu_{opt} = 0.25$, although we checked that the estimated parameter used for generating synthetic graphs is very close to~$0.6$.

For real-world and synthetic networks, the obtained value $\mu_{opt}$ can be both larger and smaller than $\mu_0 = 0.3$. Also, for synthetic networks, $\mu_0$ is close to the obtained $\mu_{opt}$. We conclude that the chosen default value is reasonable.

\begin{table*}[t]
	\caption{ILFR algorithm, default value $\mu_0 = 0.3$, standard deviation is given in the brackets}
	\label{tab:ilfr}
	\centering
	\begin{scriptsize}
		\begin{tabular}{l|cc|cc|c|cc|cc|c|cc|cc|c}
			\multicolumn{1}{c}{} & \multicolumn{5}{c|}{Rand} & \multicolumn{5}{c|}{Jaccard} & \multicolumn{5}{c}{NMI}\\
			Dataset & 
			\multicolumn{2}{c}{Default} &
			\multicolumn{2}{c}{Tuned} & $\mu_{opt}$ & 
			\multicolumn{2}{c}{Default} & \multicolumn{2}{c}{Tuned} & $\mu_{opt}$ & 
			\multicolumn{2}{c}{Default}
			& \multicolumn{2}{c}{Tuned} & $\mu_{opt}$ \\
			\hline
			\hline
			Karate &
			0.754 & (0.026) & \textbf{0.854} & (0.040) & 0.15 &
			0.507 & (0.040) & \textbf{0.741} & (0.073) & 0.05 &
			0.633 & (0.062) & 0.633 & (0.062) & 0.30 \\
			Dolphins &
			0.583 & (0.009) & \textbf{0.623} & (0.026) & 0.15 &
			0.254 & (0.018) & \textbf{0.556} & (0.000) & 0.00 &
			\textbf{0.454} & (0.019) & 0.264 & (0.000) & 1.00 \\
			Football &
			0.992 & (0.004) & 0.993 & (0.002) & 0.45 &
			0.906 & (0.033) & 0.912 & (0.020) & 0.45 &
			0.970 & (0.007) & 0.971 & (0.004) & 0.45 \\
			Political books &
			0.725 & (0.015) & \textbf{0.818} & (0.011) & 0.15 &
			0.354 & (0.038) & \textbf{0.591} & (0.026) & 0.15 &
			0.451 & (0.014) & \textbf{0.528} & (0.017) & 0.15 \\
			Political blogs &
			0.774 & (0.025) & \textbf{0.854} & (0.037) & 0.15 &
			0.569 & (0.049) & \textbf{0.728} & (0.044) & 0.15 &
			0.440 & (0.014) & \textbf{0.531} & (0.035) & 0.20 \\
			Eu-core &
			0.886 & (0.019) & \textbf{0.944} & (0.006) & 0.50 &
			0.233 & (0.028) & \textbf{0.369} & (0.024) & 0.50 &
			0.644 & (0.020) & \textbf{0.712} & (0.012) & 0.50 \\
			Cora &
			\textbf{0.978} & (0.000) & 0.977 & (0.000) & 0.05 &
			0.062 & (0.002) & \textbf{0.097} & (0.002) & 0.05 &
			\textbf{0.550} & (0.001) & 0.432 & (0.007) & 0.00 \\
			AS &
			\textbf{0.826} & (0.000) & 0.826 & (0.000) & 1.00 &
			0.021 & (0.000) & \textbf{0.183} & (0.001) & 0.00 &
			\textbf{0.444} & (0.001) & 0.420 & (0.000) & 1.00 \\
			\hline
			\hline
			LFR-0.4 &
			1.000 & (0.000) & 1.000 & (0.000) & 0.40 &
			1.000 & (0.000) & 1.000 & (0.000) & 0.40 &
			1.000 & (0.000) & 1.000 & (0.000) & 0.40 \\
			LFR-0.5 &
			1.000 & (0.000) & 1.000 & (0.000) & 0.35 &
			0.998 & (0.010) & 0.997 & (0.013) & 0.35 &
			1.000 & (0.001) & 1.000 & (0.001) & 0.35 \\
			LFR-0.6 &
			0.999 & (0.004) & 0.999 & (0.002) & 0.25 &
			0.957 & (0.084) & 0.968 & (0.057) & 0.25 &
			0.993 & (0.010) & \textbf{0.995} & (0.003) & 0.25 \\
			LFR-0.7 &
			0.972 & (0.019) & \textbf{0.981} & (0.007) & 0.35 &
			0.347 & (0.131) & 0.341 & (0.119) & 0.30 &
			0.742 & (0.058) & 0.741 & (0.064) & 0.30 \\
			\hline
		\end{tabular}
	\end{scriptsize}
\end{table*}

In rare cases, $\mu_{opt}$ for a dataset can be quite different for different similarity measures. On AS, $\mu_{opt} = 0$ for Jaccard and $\mu_{opt} = 1$ for Rand and NMI. Note that if $\mu = 0$, then the obtained algorithm tends to group all vertices in one cluster, while for $\mu = 1$ all vertices form their own clusters. Interestingly, for the Jaccard index, such a trivial partition outperforms the default algorithm. Moreover, the algorithm putting each vertex in its own cluster has close to the best performance according to the Rand index compared to all algorithms discussed in this section (both default and tuned). We conclude that AS does not have a clear community structure. 

\begin{figure*}
\centering
\begin{subfigure}{.496\textwidth}
  \centering
  \includegraphics[width=\linewidth]{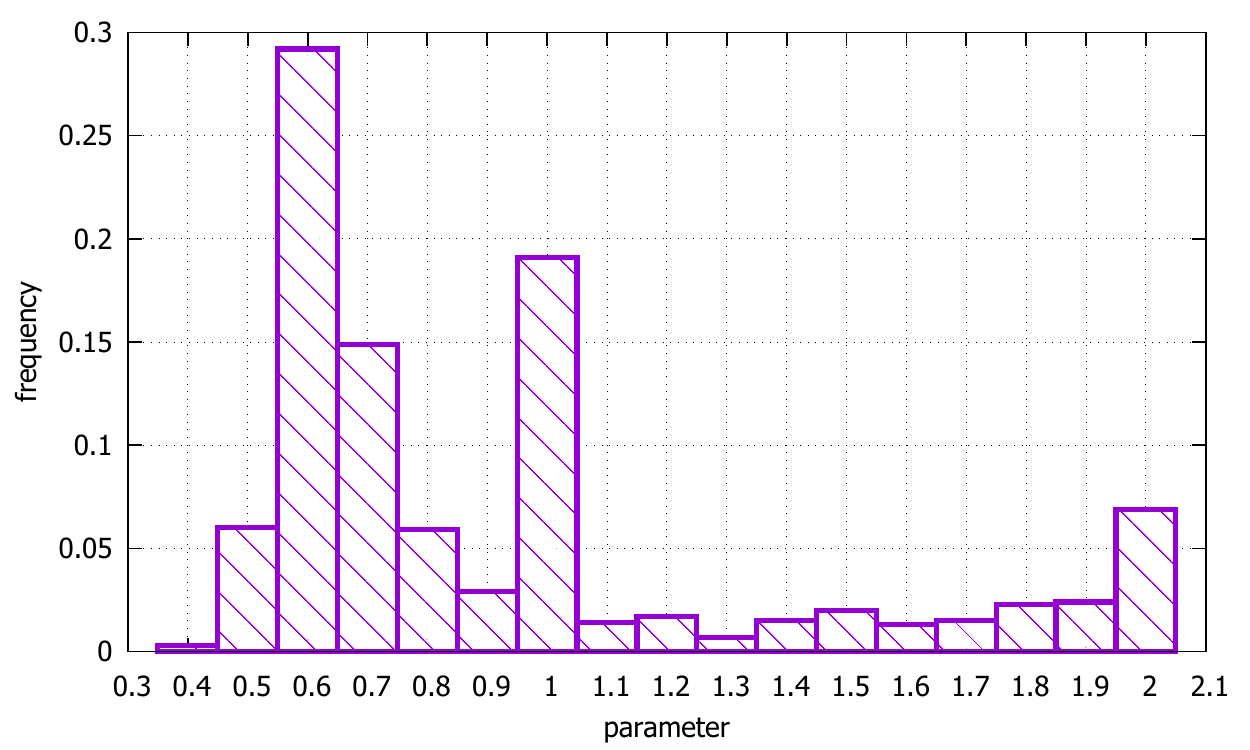}
 \caption{Karate}
 \label{fig:karate}
\end{subfigure} 
\begin{subfigure}{.496\textwidth}
  \centering
  \includegraphics[width=\linewidth]{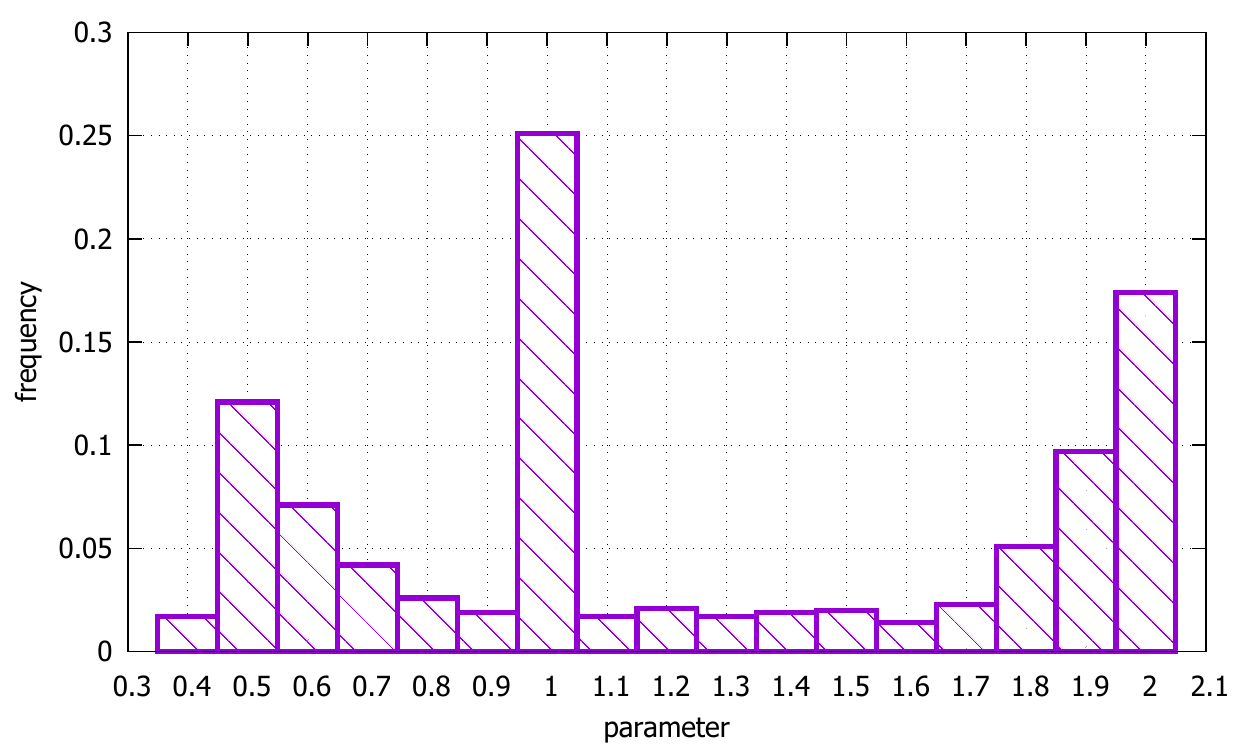}
 \caption{Dolphins}
 \label{fig:dolphins}
\end{subfigure} 
\begin{subfigure}{.496\textwidth}
  \centering
  \includegraphics[width=\linewidth]{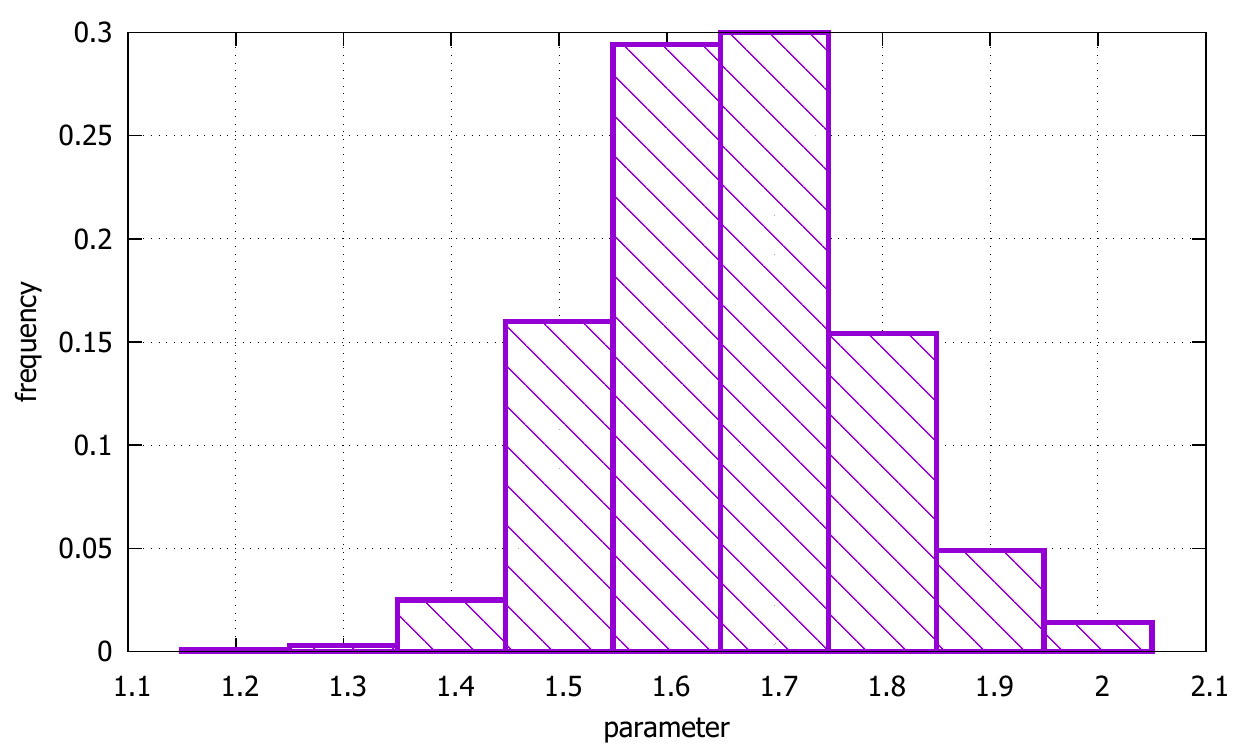}
 \caption{Football}
 \label{fig:football}
\end{subfigure} 
\begin{subfigure}{.496\textwidth}
  \centering
  \includegraphics[width=\linewidth]{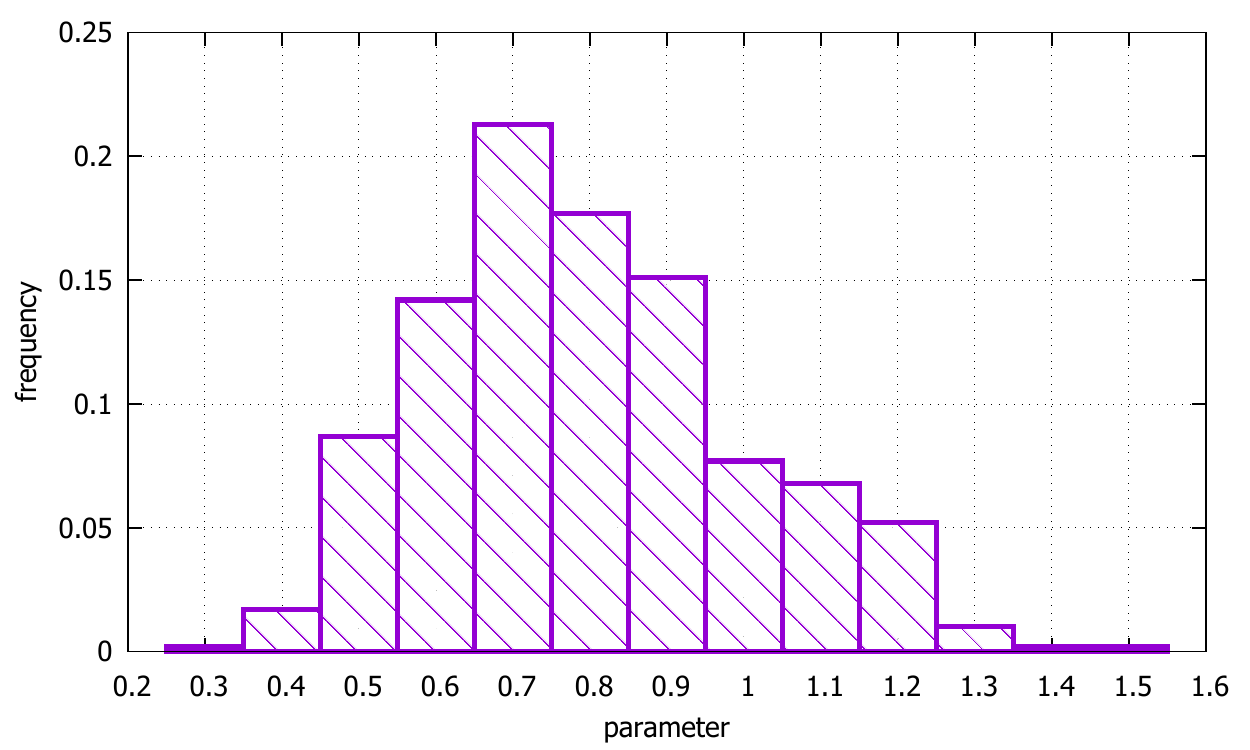}
 \caption{Political books}
 \label{fig:polbooks}
\end{subfigure} 
\begin{subfigure}{.496\textwidth}
  \centering
  \includegraphics[width=\linewidth]{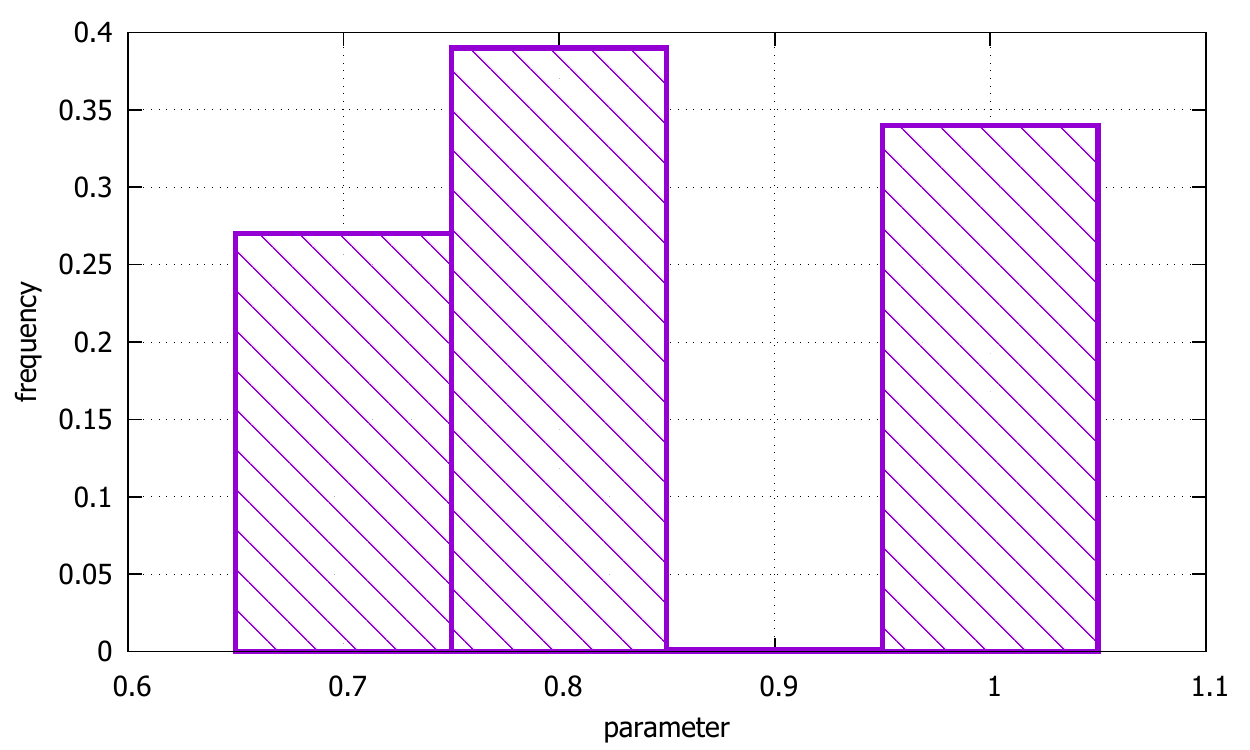}
 \caption{Political blogs}
 \label{fig:polblogs}
\end{subfigure} 
\begin{subfigure}{.496\textwidth}
  \centering
  \includegraphics[width=\linewidth]{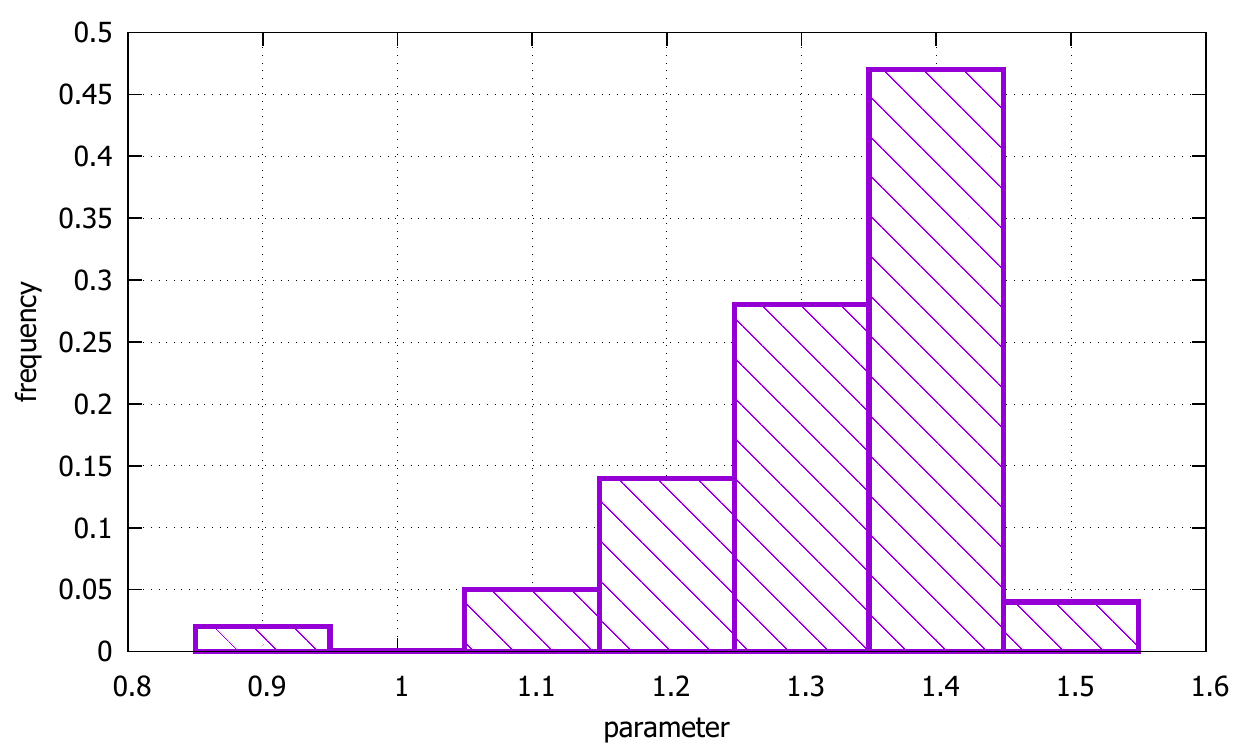}
 \caption{Eu-core}
 \label{fig:eu-cpre}
\end{subfigure}
\begin{subfigure}{.496\textwidth}
  \centering
  \includegraphics[width=\linewidth]{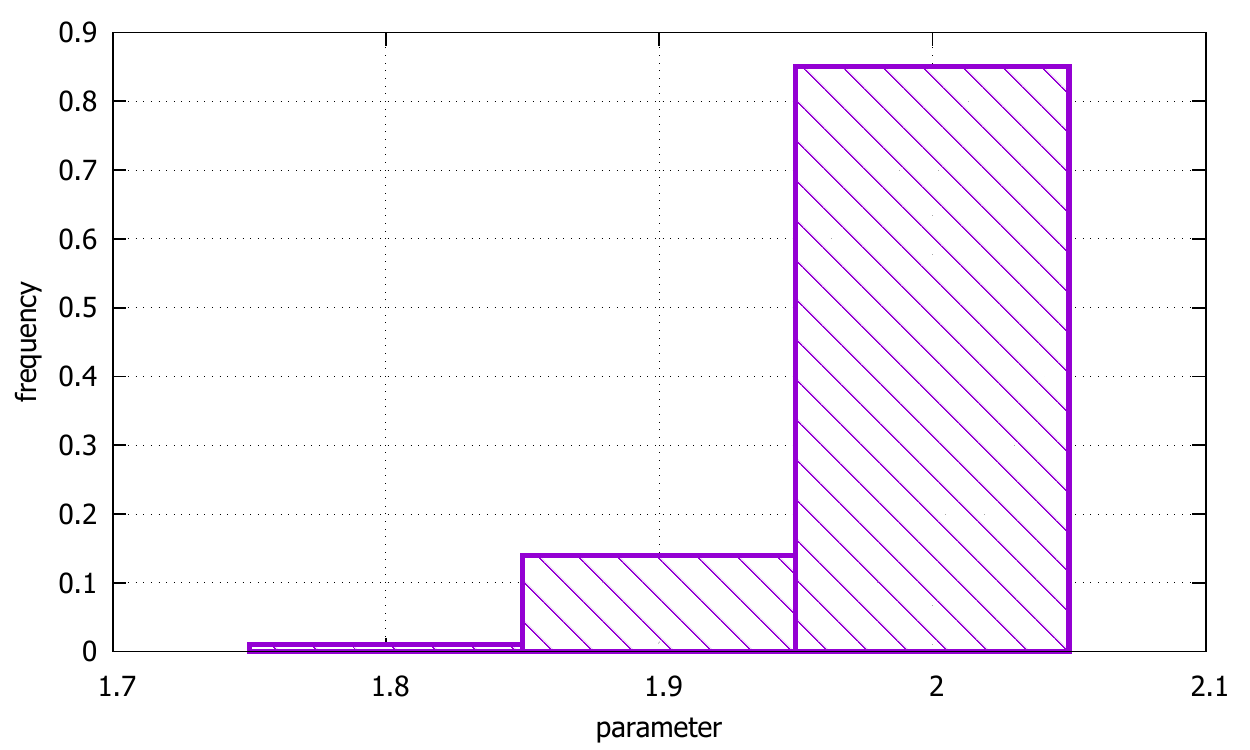}
 \caption{Cora}
 \label{fig:cora}
\end{subfigure} 
\begin{subfigure}{.496\textwidth}
  \centering
  \includegraphics[width=\linewidth]{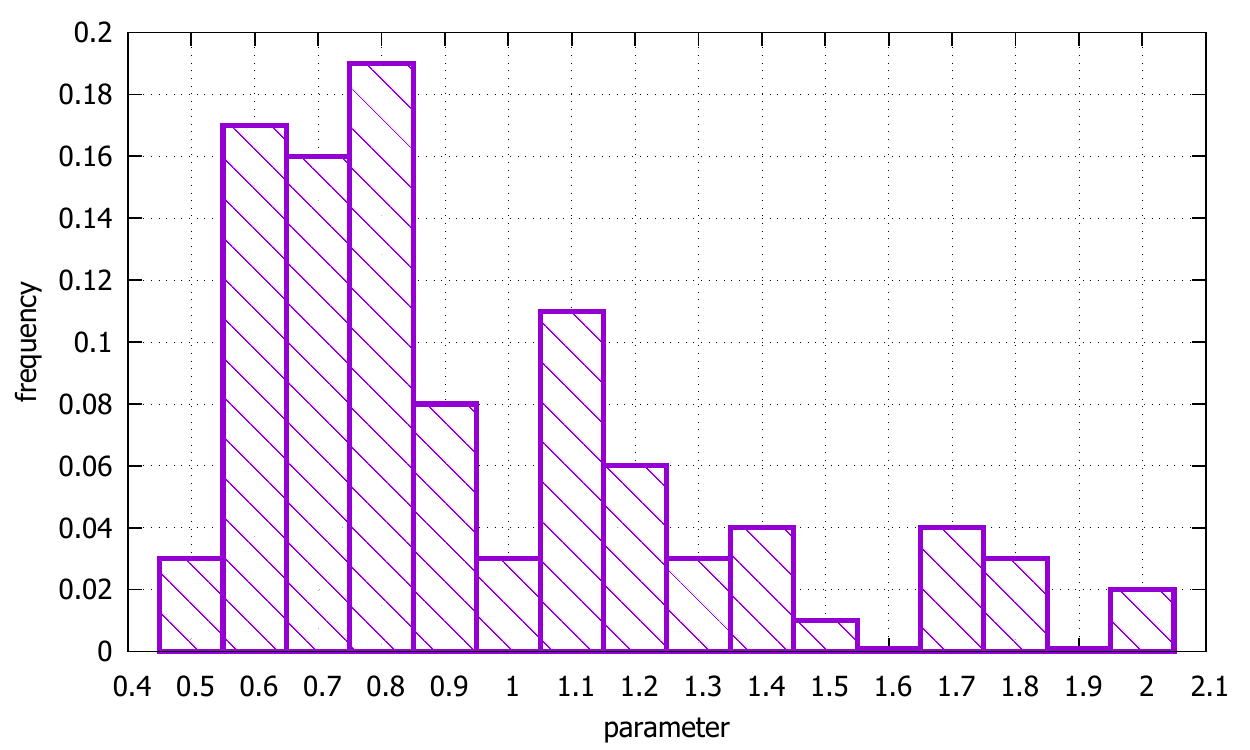}
 \caption{AS}
 \label{fig:as}
\end{subfigure}
 \caption{The distribution of $\gamma_{opt}$ for the Louvain algorithm, NMI similarity measure}
  \label{fig:hist}
\end{figure*}

\subsubsection{Stability of generated graphs}\label{sec:stability}

As discussed in Section~\ref{sec:tuning_parameters}, there are two sources of possible noise in the proposed parameter tuning procedure: 1) for small graphs the generated LFR network can be noisy, which may lead to unstable predictions of $\theta_{opt}$, 2) the randomness of $\A$ may also affect the estimate of $\theta_{opt}$ in Equation~\eqref{eq:opt_theta_honest}. The effect of the second problem can be understood using Tables~\ref{tab:louvain}-\ref{tab:ilfr}, where the standard deviations for $\theta_0$ and $\theta_{opt}$ are presented.

To analyze the effect of noise caused by the randomness in LFR graphs and to show that it is more pronounced for small datasets, we looked at the distribution of the parameters $\theta_{opt}$ obtained for different samples of generated graphs. We demonstrate this effect using the Louvain algorithm and NMI similarity measure (see Figure~\ref{fig:hist}), we take $n_{graphs} = 10^3$ for four smallest datasets and $n_{graphs} = 100$ for the other ones. Except for the AS dataset, which is noisy according to all our experiments, one can clearly see that the variance of $\gamma_{opt}$ decreases when $n$ increases. As a result, we see that for large datasets even $n_{graphs} = 1$ already provides a good estimate for $\gamma_{opt}$.

\section{Conclusion}\label{sec:conclusion}

We proposed and analyzed a surprisingly simple yet effective algorithm for hyperparameter tuning in community detection. The core idea is to generate a synthetic graph structurally similar to the observed network but with known community assignments. Using this graph, we can apply any standard black-box optimization strategy to approximately find the optimal hyperparameters and use them to cluster the original network. We empirically demonstrated that such a trick applied to several algorithms leads to significant improvements on both synthetic and real-world datasets.  Now, being able to tune parameters of any community detection algorithm, one can develop and successfully apply parametric community detection algorithms, which was not previously possible. 

\section*{Acknowledgements}

This study was funded by the Russian Foundation for Basic Research according to the research project 18-31-00207 and Russian President grant supporting leading scientific schools of the Russian Federation NSh-6760.2018.1.

\bibliographystyle{splncs04}
\bibliography{community_detection,hyperparameters} 

\end{document}